\documentstyle[12pt]{article}

\begin{document}
\begin{center}
{\LARGE \bf Evolution of Structure Functions with Jacobi Polynomial: 
Convergence and Reliability}

\vspace{0.5cm}
Sanjay K. Ghosh{\footnote {email: phys@boseinst.ernet.in}} and 
Sibaji Raha{\footnote {email: sibaji@boseinst.ernet.in}}\\  
Department of Physics, Bose Institute, \\ 93/1, A. P. C. Road, Calcutta
700 009, INDIA \\
\end{center}
                          
\vspace{0.2cm}
\begin{abstract}
The Jacobi polynomial has been advocated by many authors as a useful tool
to evolve non-singlet structure functions to higher $Q^2$. In this work,
it is found that the convergence of the polynomial 
sum is not absolute, as there is always a small fluctuation present.
Moreover, the convergence breaks down completely for large $N$.
\end{abstract}

\vspace{0.5cm}
[PACS : 12.38.-t, 12.38.Bx, 12.90.+b, 13.90.+i ] 

\vspace{0.5cm}
The structure functions are the necessary tool in our effort to understand 
the hadronic structure and strong interaction. The study of $Q^2$ evolution
of nucleon structure functions has been an important source of experimental 
information supporting Quantum Chromodynamics (QCD), which is believed to be
the fundamental theory of strong interaction. It has already been shown some 
time back 
that \cite{gluck} QCD is the only theory which can explain the gross features of
scaling violations in deep inelastic scattering (DIS). As a result a huge 
amount of effort is being put, both experimentally as well as theoretically,
to understand the nucleon structure functions for different values of
$x$ and $Q^2$.

The evolution of quark distribution with $Q^2$ is goverened by the 
Altarelli-Parisi (AP) equation \cite{alta1}. To leading order in 
$\alpha_{s}$,
the AP equation is given by,
\begin{eqnarray}
{dq(x,Q^2) \over {dt}}={\alpha_{s} \over {2\pi}} {\int_{x}}^{1}
{dy \over y} q(y,Q^2) P_{qq}({x \over y})
\label{ape}
\end{eqnarray}
where q is the quark distribution, $\alpha_{s}$ is the strong coupling, 
$t\equiv log Q^{2}$ and $P_{qq}$ is the quark splitting function which
represents the probability of a quark emitting a gluon and so becoming 
a quark with momentum reduced by a fraction ${x \over y}$.  
Eq.(\ref{ape}), an integro-differential equation, is not very easy
to solve; different methods have been proposed 
in the literature which can be grouped into three main categories.
\par
One method, already used by Altarelli, Nason and Rudolfi \cite{alta2}
is based on the assumption that, for a small variation of $t$, one can 
neglect 
the $t$ dependence of the r.h.s in eq.(\ref{ape}) and realize, in steps, the
evolution of quark distribution for a given $\delta t$. 
\par
The second method \cite{kwie} consists of expanding the quark (parton)
distribution functions into a truncated series of Chebyshev polynomial. This 
expanded form is substituted in eq.(\ref{ape}) and then the resulting
coupled differential equations are solved self-consistently.
\par
The third method relies on the premise that the moments of the structure
function (say $F_{2}$) depend only on $Q^2$. This means that one can
expand $F_{2}$ in terms of an orthonormal polynomial (usually Jacobi
polynomial) such that the $x$ dependence is carried by the polynomial
whereas the full $Q^2$ dependence is confined to the weight factors.
The variation of these weight factors with $t$ can then be extracted 
from the knowledge of the variation of moments with $t$. In the present
report we confine our attention to the use of Jacobi polynomial in
solving the evolution equation of quark distrubution functions.
\par
The Jacobi polynomial was first used by Sourlas and Parisi \cite{pari} and
later on elaborated by Barker et al. \cite{bark} and also Chyla et al.
\cite{chyla}. We will follow the prescription of ref.\cite{bark}, though
 we have found that, both \cite{pari} and \cite{bark} give similar results.
For illustrative purposes, we restrict ourselves to the valence part
of $F_{2}(x,Q^2)$.
\par
The method of orthogonal polynomials (here Jacobi Polynomial)
is based on inverting moments with the help of orthogonal polynomials.
The Jacobi polynomial is defined \cite{bark} as
\begin{eqnarray}
{{\Theta_{k}}^{\alpha,\beta}}(x) = {\sum_{j=0}}^{k} C_{k,j}(\alpha,\beta)
x^{j}
\label{jac1}
\end{eqnarray}
satisfying a weighted orthogonality
relation,
\begin{eqnarray}
{\int_{0}}^{1} dx x^{\beta} (1-x)^{\alpha} 
{{\Theta_{k}}^{\alpha,\beta}}(x) {{\Theta_{l}}^{\alpha,\beta}}(x) 
= \delta_{kl}
\label{ortho}
\end{eqnarray}
Now the structure function $F_{2}$ can be expanded as 
\begin{eqnarray}
F_{2}(x,Q^2) = x^{\beta} (1-x)^{\alpha} {\sum_{k=0}}^{\infty}
{a_{k}}^{\alpha \beta}(Q^2) {{\Theta_{k}}^{\alpha,\beta}}(x) 
\label{f2}
\end{eqnarray}
From eq.(\ref{f2}) one can write the expansion coefficients $a$ in terms
of $F_{2}$ as
\begin{eqnarray}
{a_{k}}^{\alpha \beta}(Q^2) = {\int_{0}}^{1} F_{2}(x,Q^2) 
{{\Theta_{k}}^{\alpha,\beta}}(x) dx
\label{ak1}
\end{eqnarray}
where the orthogonality relation of ${{\Theta_{k}}^{\alpha,\beta}}(x)$
(eq.\ref{ortho}) has been made use of.

Substituting eq.(\ref{jac1}) in the eq.(\ref{ak1}) one gets,
\begin{eqnarray}
{a_{k}}^{\alpha \beta}(Q^2) = {\sum_{j=0}}^{k} C_{k j}(\alpha,\beta)
\mu(j+2,Q^2)
\label{ak2}
\end{eqnarray}
where the moments $\mu$ are given by,
\begin{eqnarray}
\mu(j,Q^2) = {\int_{0}}^{1} dx x^{j-2} F_{2}(x,Q^2)
\label{moment}
\end{eqnarray}
We have used the general form of Jacobi polynomial \cite{bark,chyla},
\begin{eqnarray}
{{\Theta_{k}}^{\alpha,\beta}}(x) = {N_{k}}^{\alpha \beta}
{H_{k}}^{\alpha \beta}(x)
\end{eqnarray}
where the normalization factor is
\begin{eqnarray}
{N_{k}}^{\alpha \beta} &=& {{\Theta_{k}}^{\alpha,\beta}}(0) \nonumber \\
&=& (\beta + 1)_{k} \left[{(2k + \alpha + \beta +1) 
\Gamma (k + \alpha + \beta)} \over {\Gamma (k + \alpha + 1)
\Gamma (k + \beta +1) k!}\right]^{1 \over 2}
\label{norm}
\end{eqnarray}
with $(a)_{n} \equiv a(a+1)\cdots (a+n-1)$ and $a_{0}=1$. The expansion 
coefficients $C$ are then,
\begin{eqnarray}
C_{kj}(\alpha,\beta) = (-1)^{j} \left( {\begin{array}{c}
k \\ j \end{array}} \right) {N^{\alpha\beta}}_{k}
\frac {(k+\alpha+\beta+1)_{j}}{(\beta+1)_{j}}
\end{eqnarray}
The Jacobi Polynomial can be evaluated using eq.(\ref{jac1}) or 
the recurrence relations for the polynomial ${H_{k}}^{\alpha \beta}$ 
\cite{bark,chyla}.
\par
In the following discussion we will study the reliability and convergence
of Jacobi polynomial method for the evolution of structure functions. This
is done firstly by starting from an analytical fitted formula for valence 
part of $F_2$ at a
particular $Q^2$ and then reevaluating the $F_2$ at the same $Q^2$ using
the Jacobi polynomials. We have used three different formulae at 
$Q^2=3.5, 5$ and $15$ GeV$^2$ as given below \cite{barger,sloan,eichten},

${Q^2}_{0}=3.5$ GeV$^2$ 
\begin{eqnarray}
u(x,{Q_0}^{2})=\sqrt{x}(1-x^2)^3 (0.594+0.461(1-x^2)+0.621(1-x^2)^2) \nonumber \\
d(x,{Q_0}^{2})=\sqrt{x}(1-x^2)^3 (0.072+0.206(1-x^2)+0.621(1-x^2)^2) 
\end{eqnarray}
${Q_0}^{2}=5$ GeV$^2$ 
\begin{eqnarray}
xu_{v}(x,{Q_0}^{2})=1.78 \sqrt{x}(1-x^{1.51})^{3.5} \nonumber \\
xd_{v}(x,{Q_0}^{2})=0.67 x^{0.4} (1-x^{1.51})^{4.5} \nonumber \\
\end{eqnarray}
${Q^2}_{0}=15$ GeV$^2$ 
\begin{eqnarray}
xu_{v}(x,{Q_0}^{2})={\frac{2}{B(\alpha_{u},\beta_{u}+1)}} x^{\alpha_{u}}
(1-x)^{\beta_{u}} \nonumber \\
xd_{v}(x,{Q_0}^{2})={\frac{1}{B(\alpha_{d},\beta_{d}+1)}} x^{\alpha_{d}}
(1-x)^{\beta_{d}} 
\end{eqnarray}
where $B(\alpha,\beta)$ are the Euler beta functions, $\alpha_u = 0.588 \pm
0.020 \pm 0.05$, $\beta_u = 2.69 \pm 0.13 \pm 0.21$, $\alpha_d = 1.03 \pm
0.10 \pm 0.19$ and $\beta_d = 6.87 \pm 0.64 \pm 0.80$ \cite{sloan}.
\par  
The variation of $R \equiv \frac {{F_{2}^p}(calculated)}{{F}_{2}^p(formula)}$ 
with the $N$ (number of terms of the Jacobi polynomial) for
$Q^2=15,5$ and $3.5$ GeV$^2$ is shown in figure 1. For each $Q^2$ curves for
$x=0.05, 0.4$ and $0.75$ are plotted. We find that for $Q^2=3.5$ GeV$^2$, there
are large fluctuations in the $R$ for smaller values of $N$. The value of
$R$ becomes 1 around $N=5$ and stays the same upto a value of $N$ around 20.
But immediately after that the value of $R$ diverges. For larger values of
$x$, $R$ diverges for larger $N$. Similar features are present for 
higher values of $Q^2$ along with the additional feature that there are
small oscillations in $R$ around 1 till it diverges for $N > 20$. This 
behaviour of $R$ shows that though there is an apprent convergence
of $F_2$ with $N$, this convergence may not be as conclusive as claimed
by earlier authors \cite{pari,bark,chyla}.
\par
The unreliability of the present method of evolution of structure
functions is illustrated in figure 2. Here we have plotted the evolved 
value of only the valence part of ${F_2}^{p}$ from ${Q_0}^{2}=3.5$ GeV$^2$
to $Q^2=5$ and $15$ GeV$^2$, for $x=0.05, 0.4$ and $0.75$. Here we find
that ${F_2}^{p}$ diverges for $N \geq 12$ which is much lower than the one
observed ($N \geq 20$) in figure 1. Here again we find that the ${F_2}^{p}$
diverges earlier for lower values of $x$. The implication of these observations
is that the even the apparent convergence of $R$  does not say anything 
conclusive regarding the number of terms needed in the Jacobi Polynomial.
\par
In figures 3 and 4, we have plotted the evolved value of 
$ \Delta F={F_2}^{p}-{F_2}^{n}$ 
starting from ${Q_0}^{2} = 3.5$ and $5$ GeV$^2$ respectively.
The figures show that the $\Delta F$ gives same values for
$N \leq 14$. For $N=17$ there is a large oscillation in the value of 
$\Delta F$. The oscillations are larger for lower values of $x$
and  they decrease with increase in $x$. Furthermore, the evolved 
$\Delta F$ at $Q^2=7, 12$ and $15$ GeV$^2$ have similar values.
The above observations are valid for both ${Q_0}^{2} = 3.5$ as 
well as ${Q_0}^{2} = 5$ GeV$^2$; fluctuations being less for
${Q_0}^{2} = 5$ GeV$^2$.
\par
In conclusion we have studied the reliability and convergence of
the method of Jacobi polynomials for the evolution of structure
functions. We find that the convergence of this method is not wholly reliable
in the sense that for large $N$, the valence part of $F_2$ diverges and
there are large fluctuations in $ {F_2}^{p}-{F_2}^{n}$. Furthermore, the
final results at higher $Q^2$ are strongly dependent on the initial
fitted formula used, even for the value of $N$ where fluctuations are
not problematic. This is exemplified by the fact that even for $N$ =10, 
the final value of $F_2^p - F_2^n$ at $Q^2$ = 15 GeV$^2$ is strongly 
dependent on whether one starts from $Q^2$ = 3.5 GeV$^2$ or 5 GeV$^2$ 
(figs. 3(c) and 4(c)). Thus, the method of Jacobi Polynomials, its 
simplicity and apparent success notwithstanding, is seen to be of limited
validity in evaluating the $Q^2$ evolution of structure functions. It may
be interesting and worthwhile to explore if there are any other polynomials
which could perform better in this respect. 
\par
We are grateful to X. Song for helpful discussions and to James S.
McCarthy for helpful discussions as well as collaboration on a related
project, the results of which would be published shortly. 
SKG would like to thank the Council of Scientific and Industrial Research
(Govt. of India) for financial support.

\newpage

\centerline{FIGURE CAPTIONS}

Fig. 1. The ratio of the calculated value of the valence part
of $F_2^p$ and the corresponding value from the fitted formula
(eq.(11-13)) , starting from a given value of $Q^2$ and 
reconstructing it back at the same $Q^2$ using the Jacobi 
polynomial. Three curves in each plot correspond to x=0.05
(straight line), x=0.4 (long dash) and x=0.75 (short dash),
respectively.

Fig. 2. Evolved value of only the valence part of $F_2^p$
(a) from 3.5 GeV$^2$ to 5 GeV$^2$ and (b) from 3.5 GeV$^2$ to
15 GeV$^2$. 1, 2 and 3 in (a) and (b) correspond to x=0.05, 0.4
and 0.75 respectively.

Fig. 3. The evolved value of $F_2^p - F_2^n$ from $Q_0^2 = 3.5$
GeV$^2$ to $Q^2$ = (a) 7 GeV$^2$, (b) 12 GeV$^2$ and (c) 15
GeV$^2$. The evolution for $N$=6, 10 and 14 give the same value 
(solid line) whereas $N$=17 (dotted line) shows fluctuations
for all the cases.

Fig. 4. The evolved value of $F_2^p - F_2^n$ from $Q_0^2 = 5$
GeV$^2$ to $Q^2$ = (a) 7 GeV$^2$, (b) 12 GeV$^2$ and (c) 15
GeV$^2$. The evolution for $N$=6, 10 and 14 give the same value 
(solid line) whereas $N$=17 (dotted line) shows fluctuations
for all the cases.
\vfill
\eject
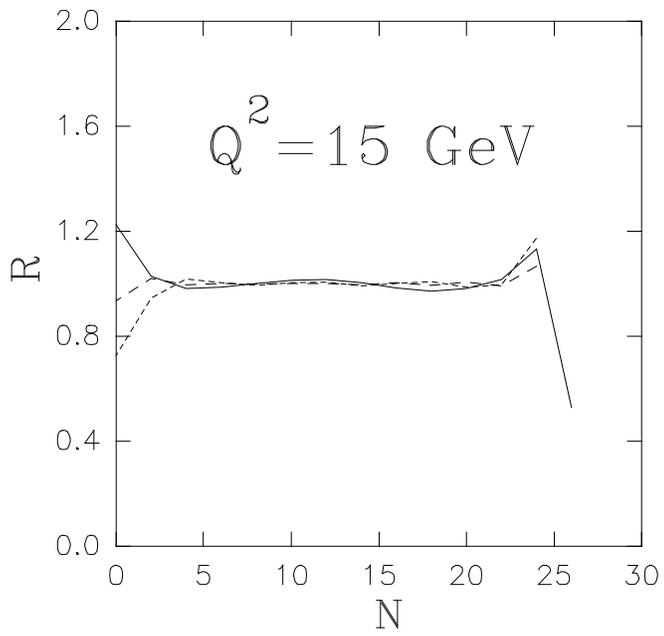
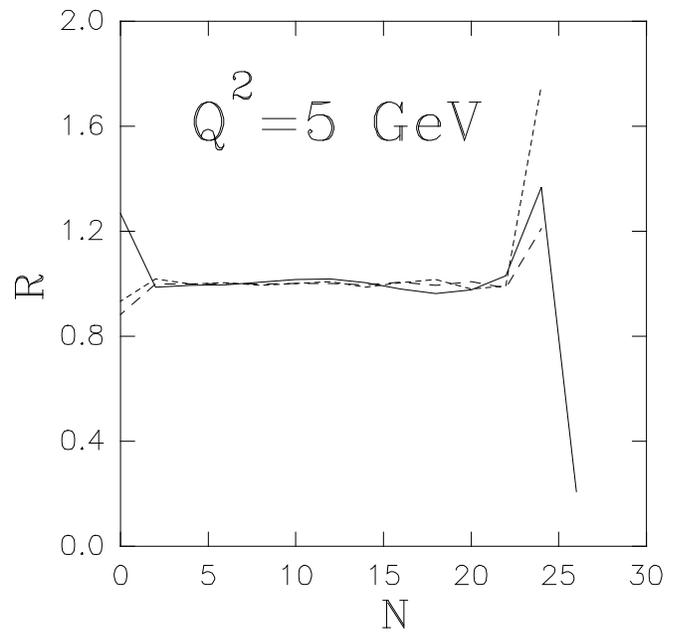
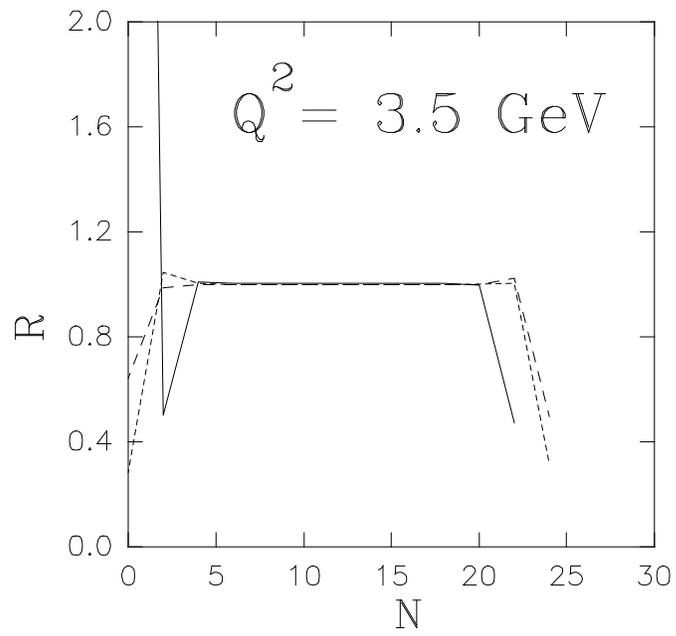
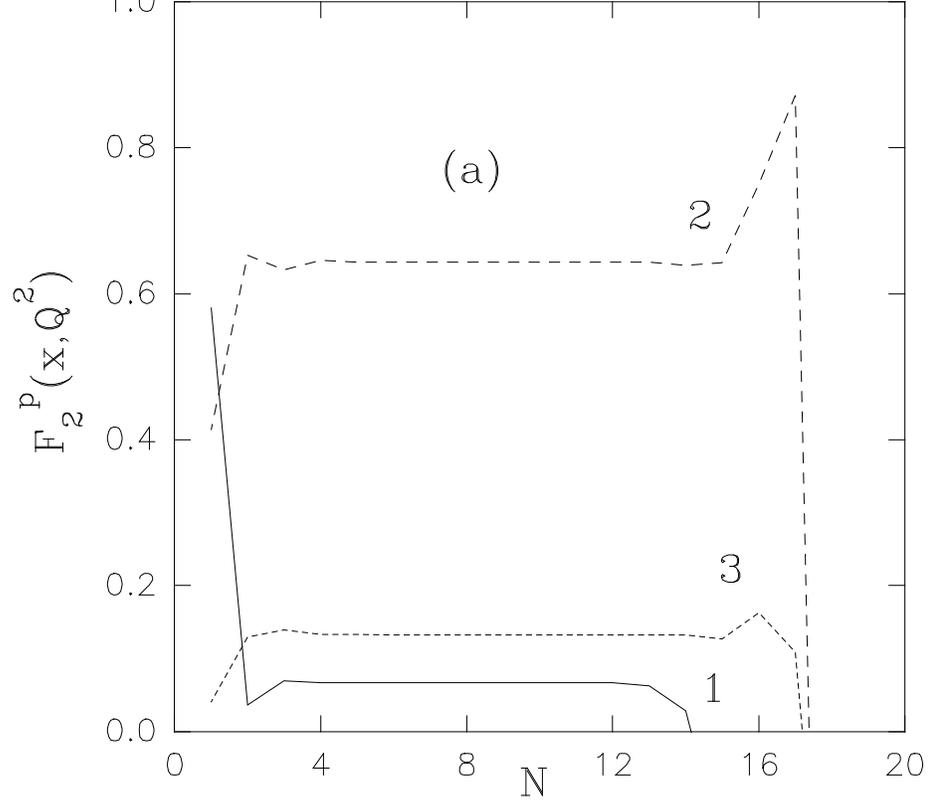
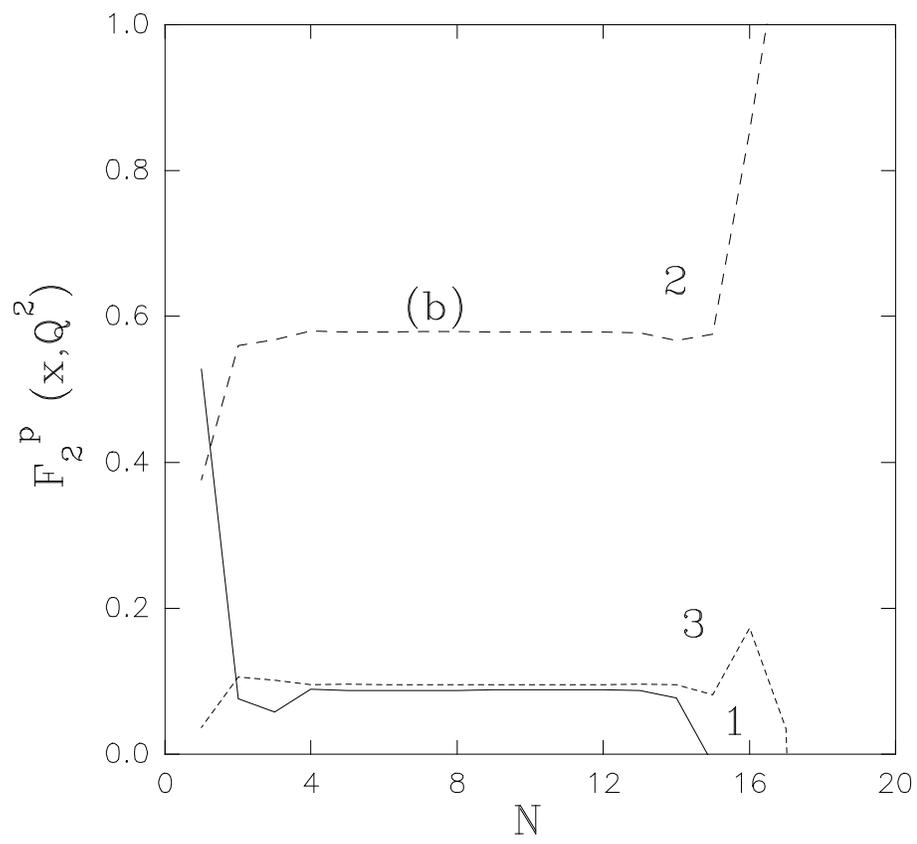
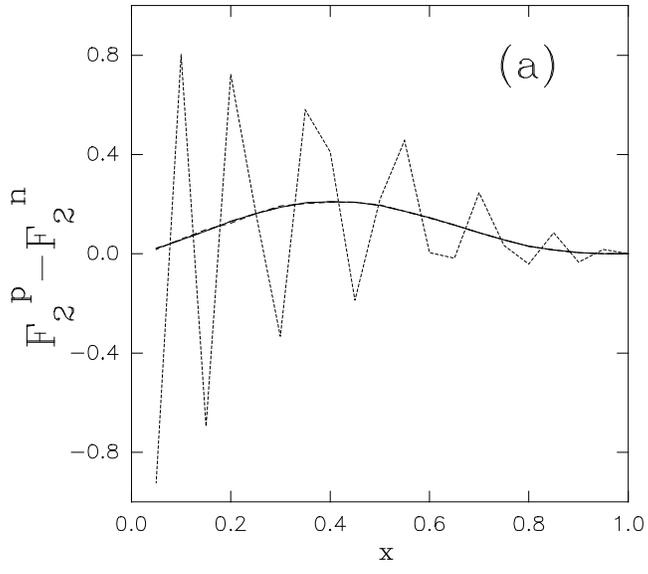
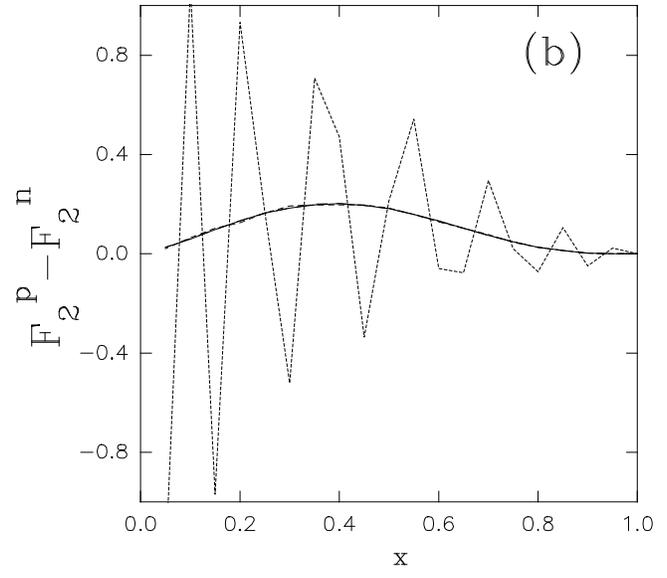
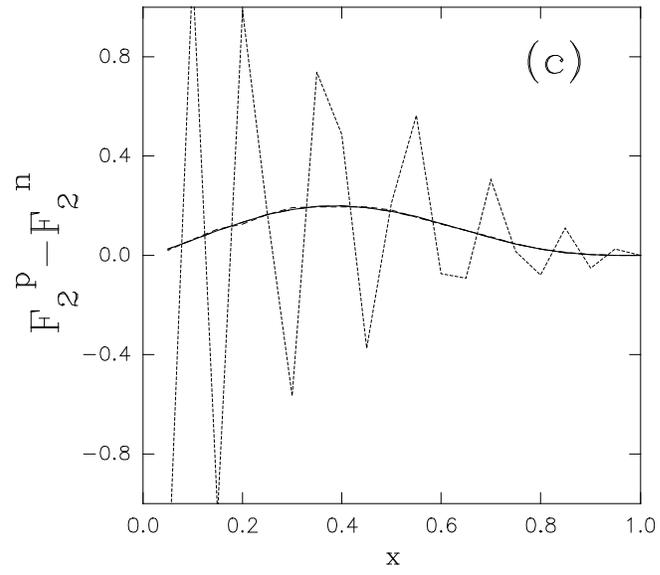
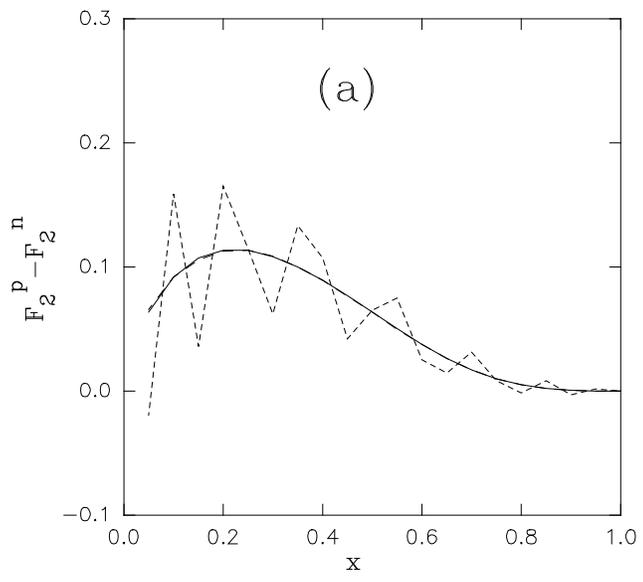
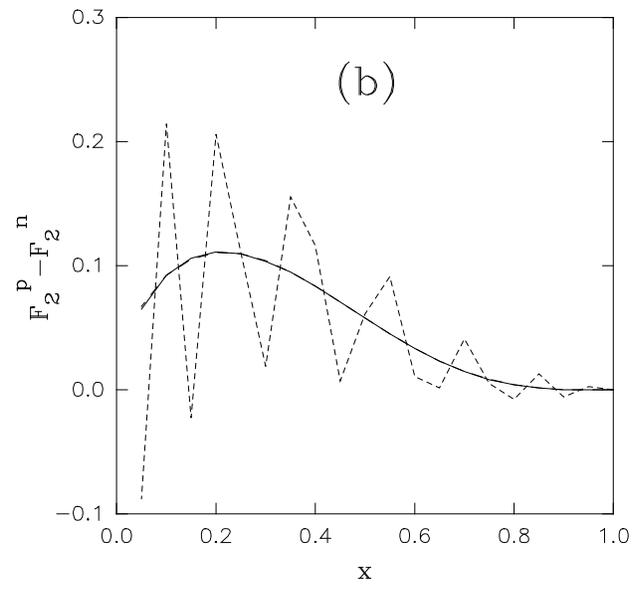
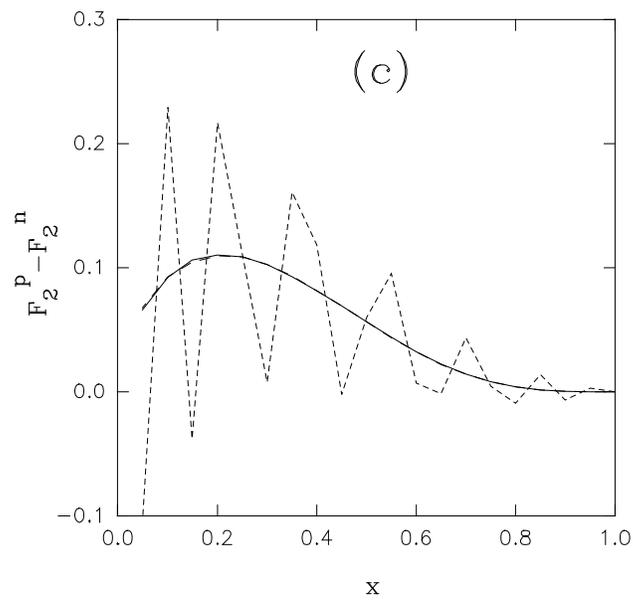
\end{document}